\documentclass[submission,copyright,creativecommons]{eptcs}
\providecommand{\event}{ICLP 2020} 
\usepackage{breakurl}             
\usepackage{underscore}           

\usepackage{mathptmx}
\usepackage{import}

\usepackage{ifthen}
\usepackage{url}
\usepackage{amssymb}
\usepackage{amsmath}
\usepackage{import}
\usepackage{xparse}
\usepackage{amsmath}
\usepackage[usenames,dvipsnames]{xcolor}
\usepackage{xspace}
\usepackage{datatool}
\usepackage{glossaries}

\providecommand\m[1]{\ensuremath{#1}\xspace}
\renewcommand{\m}[1]{\ensuremath{#1}\xspace}
\newcommand{\trval}[1]{\m{\mathbf{#1}}}

	\newcommand{\lrule}{\leftarrow}
	\newcommand{\cause}{\stackrel{c}{\lrule}}
	
	\newcommand{\ltrue}{\trval{t}}
	\newcommand{\lfalse}{\trval{f}}

	\newcommand{\voc}{\m{\Sigma}}

	\newcommand{\struct}{\m{I}}
	
	\newcommand{\I}{\m{\mathcal{I}}}

	\newcommand{\theory}{\m{\mathcal{T}}}

	\newcommand{\PP}{\m{\mathcal{P}}}

	\NewDocumentCommand\inter{g+g}{%
	  \IfNoValueTF{#1}
	    {\struct}
	    {\m{#1^{#2}}}}

	\renewcommand{\int}{\m{\mathbb{Z}}}

	\NewDocumentCommand\subs{g+g}{%
	  \IfNoValueTF{#1}
	    {\m{/}}
	    {\m{#1/ #2}}}

	\newcommand{\logicname}[1]{\textsc{#1}\xspace}
	\newcommand{\toolname}[1]{\textsc{#1}\xspace}

	\newcommand{\gringo}{\logicname{gringo}}
	\newcommand{\lparse}{\logicname{Lparse}}
	\newcommand{\smodels}{\logicname{Smodels}}

\newcommand{\ouracronym}[3]{%
	\newacronym{#1}{#2}{#3}
	\expandafter\newcommand\csname #1\endcsname{\gls{#1}\xspace}%
}
	\ouracronym{FO}{FO}{first-order logic}
	\ouracronym{PC}{PC}{propositional calculus}
	\ouracronym{MX}{MX}{Model Expansion}
	\ouracronym{MO}{MO}{Model Optimization}
	\ouracronym{ASP}{ASP}{Answer Set Programming}
	\ouracronym{TP}{TP}{Theorem Proving}
	\ouracronym{LP}{LP}{Logic Programming}
	\ouracronym{CP}{CP}{Constraint Programming}
	\ouracronym{FP}{FP}{Functional Programming}
	\ouracronym{KR}{KR}{Knowledge Representation}
	\ouracronym{CSP}{CSP}{Constraint Satisfaction Problem}
	\ouracronym{SMT}{SMT}{SAT Modulo Theories}
	\ouracronym{KBS}{KBS}{knowledge base system}
	\ouracronym{NNF}{NNF}{Negation Normal Form}
	\ouracronym{FNNF}{FNNF}{Flat Negation Normal Form}
	\ouracronym{DefNNF}{DefNNF}{Definition Negation Normal Form}
	\ouracronym{DEFNF}{DEFNF}{Definition Normal Form}
	\ouracronym{CDCL}{CDCL}{Conflict-Driven Clause-Learning}
	\ouracronym{WFS}{WFS}{Well-Founded Semantics}
	\ouracronym{LCG}{LCG}{Lazy Clause Generation}
	\ouracronym{AEL}{AEL}{Autoepistemic Logic}
	\ouracronym{OEL}{OEL}{Ordered Epistemic Logic}
	\ouracronym{AFT}{AFT}{Approximation Fixpoint Theory}

	\makeatletter
	\def\ifenv#1{
	\def\@tempa{#1}%
	\def\@ttempa{#1*}%
	\ifx\@tempa\@currenvir
	\expandafter\@firstoftwo
	\else
	\expandafter\@secondoftwo
	\fi
	}
	\makeatother

	\newcommand{\ddrule}[4]{\ensuremath{#1 \leftarrow #2 & \{#3\} & #4}}
	\newcommand{\drule}[2]{\ensuremath{#1 & \leftarrow & #2}}

	\newcommand{\darule}[4]{\ensuremath{#1 \leftarrow #2 & \{#3\} & #4}}
	\newcommand{\arule}[2]{\ensuremath{#1 \, &\leftarrow \, #2}}

	\newcommand{\LNDRule}[2]{
	\ifenv{array}
	{\drule{#1}{#2}}
	{ \ifenv{align}
		{\arule{#1}{#2}}
		{\ifenv{align*}
		{\arule{#1}{#2}}
		{ERROR: using LDRule in unsupported environment: \@currenvir}
		}
	}
	}

	\newcommand{\LDRule}[4]{
	\ifenv{array}
	{\ddrule{#1}{#2}{#3}{#4}}
	{ \ifenv{align}
		{\darule{#1}{#2}{#3}{#4}}
		{\ifenv{align*}
		{\darule{#1}{#2}{#3}{#4}}
		{ERROR: using LDRule in unsupported environment: \@currenvir}
		}
	}
	}

	\NewDocumentCommand\LRule{m+g+g+g}{%
		\IfNoValueTF{#2}%
		{#1.&}{%
		\IfNoValueTF{#3}
		{\LNDRule{#1}{#2.}}
		{\LDRule{#1}{#2.}{#3}{#4}}%
		}
	}

	\NewDocumentCommand\CLRule{m+g}{%
	\ifenv{array}
	{\cdrule{#1}{#2}}
	{ \ifenv{align}
		{\carule{#1}{#2}}
		{\ifenv{align*}
			{\carule{#1}{#2}}
			{ERROR: using CLRule in unsupported environment: \@currenvir}
		}
	}
	}

	\NewDocumentCommand\carule{m+g}{%
		\IfNoValueTF{#2}
			{\ensuremath{#1.}}
			{\ensuremath{#1 \, &\cause \, #2}}}
	\NewDocumentCommand\cdrule{m+g}{%
		\IfNoValueTF{#2}
			{\ensuremath{#1.}}
			{\ensuremath{#1 & \cause & #2}}}

	\newcommand{\algrule}[4]{
	\hbox{{#1}:}& 
	\quad #2 ~\longrightarrow~ #3 
	\hbox{~ if } #4\\
	}

	\newcommand{\AlgoRule}[4]{
	\ifenv{array}
	{\algrule{#1}{#2}{#3}{#4}}
		{ERROR: using AlgoRule in unsupported environment: \@currenvir}
	}

	\newcommand{\ignore}[1]{}

	\newboolean{nocomments}
	\setboolean{nocomments}{false}

	\newboolean{commentmargin}
	\setboolean{commentmargin}{true}

	\newcommand{\namedcomment}[3]{%
		\ifthenelse{\boolean{nocomments}}%
		{}
		{
			\ifthenelse{\boolean{commentmargin}}%
				{ {\color{#3} \marginpar{\color{#3}\sc #2}#1}  }
				{  {\color{#3} {\sc #2}: #1}  }
		}%
	}
	\newcommand{\mnamedcomment}[3]{\ifthenelse{\boolean{nocomments}}{}{{\marginpar{ \color{#3}{\sc #2}:#1}}}}

	\usepackage{soul}

\usepackage[normalem]{ulem} 
\makeatletter
\font\uwavefont=lasyb10 scaled 700
\def\spelling{\bgroup\markoverwith{\lower3.5\p@\hbox{\uwavefont\textcolor{Red}{\char58}}}\ULon}
\def\grammar{\bgroup\markoverwith{\lower3.5\p@\hbox{\uwavefont\textcolor{LimeGreen}{\char58}}}\ULon}
\def\phrasing{\bgroup\markoverwith{\lower3.5\p@\hbox{\uwavefont\textcolor{RoyalBlue}{\char58}}}\ULon}

\newcommand\remove{\bgroup\markoverwith{\textcolor{red}{\rule[0.5ex]{2pt}{0.4pt}}}\ULon}
\makeatother

\usepackage{etoolbox}

\newcommand\setcitation[2]{%
  \csdef{mycommoncitation#1}{#2}}
\newcommand\getcitation[1]{%
  \csuse{mycommoncitation#1}}

\setcitation{IDP}{WarrenBook/DeCatBBD14}
\setcitation{idp}{WarrenBook/DeCatBBD14}
\setcitation{fodot}{tocl/DeneckerT08}
\setcitation{foid}{tocl/DeneckerT08}
\setcitation{FOID}{tocl/DeneckerT08}
\setcitation{cplogic}{journal/tplp/VennekensDB10}
\setcitation{CPlogic}{journal/tplp/VennekensDB10}
\setcitation{CPLogic}{journal/tplp/VennekensDB10}
\setcitation{CP}{fai/Rossi06}
\setcitation{cp}{fai/Rossi06}
\setcitation{EZCSP}{lpnmr/Balduccini11}
\setcitation{KR}{Baral:2003}
\setcitation{ASPComp2}{lpnmr/DeneckerVBGT09}
\setcitation{ASPComp3}{journals/tplp/CalimeriIR14}
\setcitation{ASPComp4}{conf/lpnmr/AlvianoCCDDIKKOPPRRSSSWX13}
\setcitation{ASPComp5}{journals/ai/CalimeriGMR16}
\setcitation{ASPComp6}{jair/GebserMR17}
\setcitation{ASPComp7}{tplp/GebserMR20}
\setcitation{CPSupport}{ictai/DeCat13}
\setcitation{CPsupport}{ictai/DeCat13}
\setcitation{functionDetection}{iclp/DeCatB13}
\setcitation{FunctionDetection}{iclp/DeCatB13}
\setcitation{fodot2asp}{corr/DeneckerLTV19} 
\setcitation{Tarskian}{corr/DeneckerLTV19} 
\setcitation{TarskianSemanticsASP}{corr/DeneckerLTV19} 
\setcitation{Inca}{iclp/DrescherW12}
\setcitation{csp2asp}{ijcai/DrescherW11}
\setcitation{DPLLT}{cav/GanzingerHNOT04}
\setcitation{AspInPractice}{synthesis/2012Gebser}
\setcitation{ASPInPractice}{synthesis/2012Gebser}
\setcitation{clasp}{ai/GebserKS12}
\setcitation{oclingo}{kr/GebserGKOSS12}
\setcitation{clingo}{iclp/GebserKKOSW16}
\setcitation{gringo}{lpnmr/GebserST07}
\setcitation{cmodels}{aaai/GiunchigliaLM04}
\setcitation{inputster}{tplp/Jansen13}
\setcitation{DLV}{tocl/LeonePFEGPS06}
\setcitation{LearningPaper}{TPLP/BruynoogheBBDDJLRDV} 
\setcitation{clog}{iclp/BogaertsVDV14} 
\setcitation{foc}{iclp/BogaertsVDV14}
\setcitation{FOC}{iclp/BogaertsVDV14}
\setcitation{inferenceClog}{ecai/BogaertsVDV14}
\setcitation{examplesClog}{nmr/BogaertsVDV14b} 
\setcitation{AFT}{DeneckerMT00}
\setcitation{KBS}{iclp/DeneckerV08}
\setcitation{KBS-invitedtalk}{jelia/Denecker16}
\setcitation{KBPE}{inap/DePooterWD11}
\setcitation{lazyGrounding}{jair/CatDBS15} 
\setcitation{LazyGrounding}{jair/CatDBS15} 
\setcitation{lazygrounding}{jair/CatDBS15} 
\setcitation{lazygroundingASP}{ijcai/BogaertsW18} 
\setcitation{justifications}{lpnmr/DeneckerBS15} 
\setcitation{justificationsAlpha}{ijcai/BogaertsW18} 
\setcitation{ASP}{marek99stable}
\setcitation{satid}{sat/MarienWDB08}
\setcitation{lazyclausegeneration}{constraints/OhrimenkoSC09}
\setcitation{FP}{ACMCS/Hudak89}
\setcitation{GroundingWithBounds}{jair/WittocxMD10}
\setcitation{GroundWithBounds}{jair/WittocxMD10}
\setcitation{SAT}{faia/SilvaLM09}
\setcitation{HandbookOfSAT}{faia/2009-185}
\setcitation{LTC}{iclp/Bogaerts14}
\setcitation{SPSAT}{ictai/DevriendtBMDD12}
\setcitation{BreakID}{sat/DevriendtBBD16}
\setcitation{breakid}{sat/DevriendtBBD16}
\setcitation{LCG}{stuckeyLCG}
\setcitation{MiniZinc}{conf/cp/NethercoteSBBDT07}
\setcitation{minizinc}{conf/cp/NethercoteSBBDT07}
\setcitation{amadini}{cpaior/AmadiniGM13}
\setcitation{bootstrapping}{ngc/BogaertsJDJBD16}
\setcitation{Bootstrapping}{ngc/BogaertsJDJBD16}
\setcitation{GroundedFixpoints}{ai/BogaertsVD15}
\setcitation{PartialGroundedFixpoints}{ijcai/BogaertsVD15}
\setcitation{LogicBlox}{datalog/GreenAK12}
\setcitation{proB}{journals/sttt/LeuschelB08}
\setcitation{NaturalInductions}{KR/DeneckerV14} 
\setcitation{LP}{jacm/EmdenK76}
\setcitation{SMT}{faia/BarrettSST09}
\setcitation{AF}{ai/Dung95}
\setcitation{ADF}{kr/BrewkaW10}
\setcitation{af}{ai/Dung95}
\setcitation{adf}{kr/BrewkaW10}
\setcitation{ADFRevisited}{ijcai/BrewkaSEWW13}
\setcitation{adfrevisited}{ijcai/BrewkaSEWW13}
\setcitation{DefaultLogic}{ai/Reiter80}
\setcitation{DL}{ai/Reiter80}
\setcitation{AEL}{mo85}
\setcitation{minisat}{sat/EenS03}
\setcitation{completion}{adbt/Clark78}
\setcitation{ClarkCompletion}{adbt/Clark78}
\setcitation{wasp}{lpnmr/AlvianoDFLR13}
\setcitation{minisatid}{ictai/DeCat13}
\setcitation{lcg}{stuckeyLCG}
\setcitation{CEGAR}{jacm/ClarkeGJLV03}
\setcitation{cegar}{jacm/ClarkeGJLV03}
\setcitation{CuttingPlane}{or/DantzigFJ54}
\setcitation{kodkod}{tacas/TorlakJ07}
\setcitation{cdcl}{Marques-SilvaS99}
\setcitation{CDCL}{Marques-SilvaS99}
\setcitation{1UIP}{iccad/ZhangMMM01}
\setcitation{relevance}{ijcai/JansenBDJD16}
\setcitation{relevance-implementation}{aspocp/JansenBDJD16}
\setcitation{WFS}{GelderRS91}
\setcitation{wfs}{GelderRS91}
\setcitation{UnfoundedSet}{GelderRS91}
\setcitation{UFS}{GelderRS91}
\setcitation{stablesemantics}{iclp/GelfondL88}
\setcitation{StableSemantics}{iclp/GelfondL88}
\setcitation{shatter}{Shatter}
\setcitation{sbass}{drtiwa11a}
\setcitation{lparsemanual}{url:lparsemanual}
\setcitation{AIC}{ppdp/FlescaGZ04}
\setcitation{templates}{tplp/DassevilleHJD15}
\setcitation{templates2}{iclp/DassevilleHBJD16}
\setcitation{sat-to-sat}{aaai/JanhunenTT16}
\setcitation{sat-to-sat-qbf}{bnp/BogaertsJT16}
\setcitation{sat-to-sat-QBF}{bnp/BogaertsJT16}
\setcitation{sat-to-sat-SO}{kr/BogaertsJT16}
\setcitation{XSB}{SwiW12}
\setcitation{KCmap}{jair/DarwicheM02}
\setcitation{TLA}{DBLP:books/aw/Lamport2002}
\setcitation{EventB}{BookAbrial2010}
\setcitation{MX}{MitchellT05}
\setcitation{MIP}{Sierksma96}
\setcitation{perefectmodel}{minker88/Przymusinski88}
\setcitation{SafeInductions}{ijcai/BogaertsVD17}
\setcitation{AIC}{ppdp/FlescaGZ04}
\setcitation{aic}{ppdp/FlescaGZ04}
\setcitation{alpha}{lpnmr/Weinzierl17}
\setcitation{omiga}{jelia/Dao-TranEFWW12}
\setcitation{gasp}{fuin/PaluDPR09}
\setcitation{asperix}{lpnmr/LefevreN09a}
\setcitation{CTL}{lop/ClarkeE81}
\setcitation{AFT-AIC}{ai/BogaertsC18}
\setcitation{UltimateApproximator}{DeneckerMT04}
\setcitation{KripkeKleene}{Fitting85}
\setcitation{AFT-HO}{corr/CharalambidisRS18} 
\setcitation{HereThere}{Heyting30}
\setcitation{dAEL}{ijcai/HertumCBD16}
\setcitation{SDD}{ijcai/Darwiche11}
\setcitation{HEX}{ijcai/EiterIST05}
\setcitation{wADF}{aaai/BrewkaSWW18}
\setcitation{wADFfix}{corr/BrewkaSWW18}
\setcitation{TransitionSystems}{jacm/NieuwenhuisOT06}
\setcitation{galliwasp}{lopstr/MarpleG12}
\setcitation{GalliWasp}{lopstr/MarpleG12}
\setcitation{clingcon}{tplp/BanbaraKOS17}
\setcitation{lp2sat}{birthday/JanhunenN11}
\setcitation{lp2mip}{LIU12}
\setcitation{lp2diff}{lpnmr/JanhunenNS09}
\setcitation{lp2acyc}{ecai/GebserJR14}
\setcitation{PB}{faia/RousselM09}
\setcitation{pb}{faia/RousselM09}
\setcitation{CuttingPlanes}{dam/CookCT87}
\setcitation{RoundingSAT}{ijcai/ElffersN18}
\setcitation{PRS}{aaai/DixonG02}
\setcitation{sat4j}{jsat/BerreP10}
\setcitation{SAT4J}{jsat/BerreP10}
\setcitation{mingo}{LIU12}
\setcitation{pbmodels}{lpnmr/LiuT05}
\setcitation{HEF-LP}{lpnmr/GebserLL07}
\setcitation{HCF-LP}{amai/Ben-EliyahuD94}
\setcitation{aspcore2}{tplp/CalimeriFGIKKLM20}
\setcitation{AspCore2}{tplp/CalimeriFGIKKLM20}
\setcitation{}{}
\setcitation{}{}
\setcitation{}{}
\setcitation{}{}
\setcitation{}{}
\setcitation{}{}
\setcitation{}{}
\setcitation{}{}
\setcitation{}{}
\setcitation{}{}
\setcitation{}{}
\setcitation{}{}
\setcitation{}{}
\setcitation{}{}
\setcitation{}{}
\setcitation{}{}
\setcitation{}{}
\setcitation{}{}
\setcitation{}{}
\setcitation{}{}
\setcitation{}{}
\setcitation{}{}
  
\newcommand\refto[1]{%
      \ifcsname mycommoncitation#1\endcsname%
      \getcitation{#1}%
      \else%
      #1%
      \fi%
      }
      
\newcommand\mycite[1]{%
      \ifcsname mycommoncitation#1\endcsname%
   \cite{\getcitation{#1}}%
  \else%
    \cite{#1}%
  \fi%
}	
  
\ignore{

}
\usepackage{tikz}
\usetikzlibrary{plotmarks}
\usepackage[mode=buildnew]{standalone}
\usepackage{caption}
\usepackage{subcaption}
\usepackage{filecontents}
\usepackage{pgfplots}
\usepackage{xcolor}
\newcommand\gurobi{\toolname{gurobi}}

\newcommand\lptopb{\toolname{lp2pb}}
\newcommand\lpshift{\toolname{lpshift}}
\newcommand\lptonormal{\toolname{lp2normal}}
\newcommand\lptolp{\toolname{lp2lp2}}
\newcommand\lptosat{\toolname{lp2sat}}
\newcommand\dlvtwo{\toolname{DLV2}}
\newcommand\clasp{\toolname{clasp}}
\newcommand\pbmodels{\toolname{pbmodels}}
\newcommand\mingo{\toolname{mingo}}
\newcommand\roundingsat{\toolname{RoundingSAT}}
\newcommand\head{\m{\mathrm{head}}}
\newcommand\body{\m{\mathrm{body}}}
\renewcommand\voc{\m{\sigma}}

\newcommand\eg{\emph{e.g.}\ }

\newcommand\bcmdtab{\noindent\bgroup\tabcolsep=0pt%
  \begin{tabular}{@{}p{10pc}@{}p{20pc}@{}}}
\newcommand\ecmdtab{\end{tabular}\egroup}

\title{\lptopb: Translating Answer Set Programs into Pseudo-Boolean Theories}
\author{Wolf De Wulf\qquad\qquad Bart Bogaerts
\institute{Vrije Universiteit Brussel\\ Brussels, Belgium}
\email{firstname.lastname@vub.be}
}
\def\titlerunning{\lptopb: Translating Answer Set Programs into Pseudo-Boolean Theories}
\def\authorrunning{W. De Wulf \& B. Bogaerts}

\newtheorem{theorem}{Theorem}
\newtheorem{remark}[theorem]{Remark}
\newtheorem{proposition}[theorem]{Proposition}

\begin{document}

\label{firstpage}

\maketitle
 
  \begin{abstract}
  Answer set programming (ASP) is a well-established knowledge representation
  formalism. Most ASP solvers are based on (extensions of) technology from Boolean satisfiability solving. 
  While these solvers have shown to be very successful in many practical applications, their strength is limited by their underlying proof system, resolution. 
  In this paper, we present a new tool \lptopb that translates ASP programs into pseudo-Boolean theories, for which solvers based on the (stronger) cutting plane proof system exist. 
  We evaluate our tool, and the potential of cutting-plane--based solving for ASP on traditional ASP benchmarks as well as benchmarks from pseudo-Boolean solving. Our results are mixed: overall, traditional ASP solvers still outperform our translational approach, but several benchmark families are identified where the balance shifts the other way, thereby suggesting that further investigation into a stronger proof system for ASP is valuable.
  \end{abstract}



\section{Introduction}

Answer set programming (ASP) is a well-established knowledge representation formalism that grew from the observation that stable models \cite{iclp/GelfondL88}
of a logic program can be used to encode search problems
\cite{marek99stable,Niemela99,iclp/Lifschitz99}.
ASP is rapidly gaining adoption, with applications in domains such as 
decision support for the Space Shuttle \cite{padl/NogueiraBGWB01},
product configuration \cite{paper:tiihonen:2003},
phylogenetic inference
\cite{tplp/KoponenOJS15,jar/BrooksEEMR07},
knowledge management \cite{lpnmr/GrassoILR09},
e-Tourism \cite{fuin/RiccaDGIIML10},
biology \cite{tplp/GebserSTV11}, 
robotics \cite{lpnmr/AndresRSS15},
and machine learning \cite{jagerinypeco15a,TPLP/BruynoogheBBDDJLRDV}.


The success of ASP can, to a large extend, be explained by two factors. 
The first factor is a rich, first-order  language, ASP-Core2~\mycite{AspCore2}, to express knowledge in, with an easy-to-understand modeling methodology known as generate-define-and-test. 
The second factor is the availability of a large number of reliable tools  --- grounders \cite{\refto{gringo},\refto{DLV}} and solvers \cite{\refto{clasp},\refto{wasp},\refto{minisatid}} --- that allow to efficiently compute stable models of a given logic program.  

Throughout its history, ASP has always benefited from progress in other domains of combinatorial search. 
For instance, the addition of conflict-driven clause learning (CDCL) \mycite{cdcl} to Boolean satisfiability (SAT) solvers is often recognized as one of the most important leaps forward in SAT solving; this technique was very quickly adopted in ASP. 
In fact, the relation goes two ways, \clasp, a native ASP solver has long been one of the best performing SAT solvers. 
A recent example of such  positive reinforcement between domains is found in recent constraint ASP systems \cite{\refto{clingcon}}, which use techniques from SAT modulo theories \mycite{SMT} and from constraint programming \mycite{CP} -- in particular, lazy clause generation \mycite{lcg}. 

Next to native solvers, also various ASP tools are available based on \emph{translations} to other formalisms: to SAT \mycite{lp2sat}, to difference logic \mycite{lp2diff}, to mixed integer programming \mycite{lp2mip}, and to SAT modulo acyclicity \mycite{lp2acyc}. 
Our current work fits in this line, by translating answer set programs into (linear) pseudo-Boolean (PB) constraints \mycite{PB}. 

Most modern ASP solvers are built on conflict-driven clause learning and thus on the resolution proof system. 
This also holds for ASP solvers with native support for aggregates, which typically employ lazy clause generation techniques, essentially compiling their theory lazily into clauses and henceforth relying on the underlying CDCL solver. 
The advantage of building on CDCL technology is that this has been researched intensely, and the simplicity of using only clauses allows for highly optimized implementations, resulting in efficient, well-engineered solvers. 
The disadvantage is that the resolution proof system is known to be weak; for several very simple problems, resolution proofs are exponentially large; the most notorious such problem is the pigeon hole problem. 
In practice this means that modern CDCL solvers can, for instance, not solve the problem ``do 15 pigeons fit in 14 holes?''
One way to avoid this limitation, is using symmetry exploitation methods, which are well-researched in SAT \cite{Shatter,sat/DevriendtBBD16,sat/DevriendtBB17,DBLP:conf/nfm/MetinBK19}, and have also been ported to ASP \cite{drtiwa11a,aspocp/DevriendtB16}, but the detection of symmetries is often very brittle, \eg, adding redundant constraints often removes symmetries. Another option is using a stronger proof system, such as the \emph{cutting planes} proof system \mycite{CuttingPlanes}. 
This proof system works on linear constraints over the integers, or, when restricted to $0-1$ variables, on so-called pseudo-Boolean constraints. 
Recent research in the field of pseudo-Boolean solving has resulted in a new and efficient solver, \roundingsat \mycite{RoundingSAT}, that builds on previous work to integrate conflict-driven search with the cutting plane proof system  \mycite{\refto{PRS},tcad/ChaiK05,\refto{jsat/SheiniS06},jsat/ManquinhoS06,\refto{SAT4J}}.
This recent improvement in psuedo-Boolean solving triggers the question whether answer set programming could also benefit from these techniques.

The main contribution of this paper is the introduction and experimental validation of a new tool \lptopb that translates ground logic programs into pseudo-Boolean theories. 
This tool is valuable both for the ASP community and for the PB community. For ASP, it enables the use of an extra class of solvers. Furthermore, since we translate into the well-accepted OPB standard format for pseudo-Boolean problems\footnote{See \url{http://www.cril.univ-artois.fr/PB10/format.pdf}.}, compatibility of  future pseudo-Boolean solvers is obtained for free, allowing us to quickly test the potential of novel PB solving techniques for logic programming. Additionally, the OPB format is supported by important industrial tools such as \gurobi \mycite{\refto{gurobi}}.
For the PB community, this tool provides access to a new set of applications and benchmarks. Additionally, it establishes answer set programming as \emph{a modelling language for PB solvers}, thereby bypassing the need to write by hand a program that generates benchmarks for every new class of benchmarks considered.

We experimentally validate 
\lptopb on two classes of benchmarks. 
On novel ASP models of four benchmark families where the difference between cutting planes and resolution is known to be essential, our approach, unsurprisingly, outperforms traditional ASP solvers. 
On benchmarks from the latest ASP competition, we find that overall, traditional ASP solvers are still more efficient, but several benchmark families (mainly optimization problems) are found where the cutting plane proof system pays off. 
    
The rest of this paper is structured as follows. 
In Section~\ref{sec:prelims}, we introduce preliminaries on ASP and pseudo-Boolean constraints. 
In Sections~\ref{sec:translation} and~\ref{sec:implementation}, we discuss our translation and its implementation respectively. 
Section~\ref{sec:experiments} contains our experiments. In Section~\ref{sec:related}, we discuss some closely related work and we conclude in Section~\ref{sec:conclusion}.

\section{Preliminaries}\label{sec:prelims}

A \emph{vocabulary} is a set of symbols, also called
\emph{atoms}; vocabularies are denoted by $\sigma,\tau$. A \emph{literal}
is an atom $p$ or its negation $\bar p$.
An interpretation $\struct$ of a vocabulary \voc is a subset of $\voc$. 
We use the truth values true ($\ltrue$) and false ($\lfalse$) and will identify $\ltrue$ with $1$ and $\lfalse$ with $0$, as is common in pseudo-Boolean theories. 
The truth value of an atom $p\in \sigma$ in an interpretation \struct  (denoted $p^\struct$) is $1$ if $p\in \struct$ and $0$ otherwise. 
The truth value of literals, conjunctions of literals, and clauses (disjunctions of literal) are defined as usual. 

\paragraph{Pseudo-Boolean Constraints}
A (linear) pseudo-Boolean constraint over $\voc$ is a linear constraint with variables from \voc, i.e., an expression of the form 
\begin{equation}\label{eq:pb}\sum_i w_i x_i\sim b\end{equation}
with $w_i,b\in\mathbb{Z}$, $x_i\in \voc$, and $\sim$ one of $<$, $>$, $\leq$, $\geq$, and $=$. 
The value of a $\sum_{i=1}^n w_i x_i$ in \struct is, as usual, defined as 
$\sum_{i=1}^n w_i x_i^\struct$. 
A pseudo-Boolean constraint of the form \eqref{eq:pb} is satisfied in \struct if $\sum_{i=1}^n w_i x_i^\struct \sim b$. 
A pseudo-Boolean theory is a set of pseudo-Boolean constraints.  
A \emph{model} of a pseudo-Boolean theory $\theory$ is an interpretation \struct such that all constraints are satisfied in \struct. 
%

\paragraph{Logic Programming}
A \emph{normal logic program} $\PP$ over vocabulary $\voc$ is a set of \emph{rules} $r$ of form 
\begin{equation}\label{eq:rule} 
h \lrule a_1\land \dots \land a_n \land \overline{b_1} \land \dots \land  \overline{b_m}.\end{equation}
where $h$, the $a_i$'s, and $b_i$'s are atoms in $\voc$. We call $h$ the \emph{head} of $r$, denoted $\head(r)$, and 
$a_1\land \dots \land a_n \land \overline{b_1}\land \dots \land \overline{b_m}$ the \emph{body} of $r$, denoted $\body(r)$.
If $n=m=0$, we simply write $h$.

\begin{remark}
 We use the notation $\overline{p}$  for the negation of $p$. 
In the context of logic programming, the type of negation used here is often referred to as ``negation as failure'' or ``default negation'', referring to the fact that in ``good'' models of logic programs (called stable models below), an atom $p$ is false by default: it is false unless there is a rule that can derive it. 
In this work, there is no need to distinguish between different types of negation (indeed, all definitions such as when an interpretation satisfies a literal remain valid) and for uniformity and brevity we thus use the notation $\overline{p}$ which is standard in pseudo-Boolean solving throughout the paper.
\end{remark}

An interpretation $\struct$ is a \emph{model} of a logic program \PP if, for all rules $r$ in
\PP, whenever $\body(r)$ is satisfied by $I$, so is $\head(r)$.  The
\emph{reduct} of \PP with respect to $I$, denoted $\PP^I$, is the program that consists of rules 
$
h \lrule a_1\land \dots \land a_n $
for all rules of the form \eqref{eq:rule} in \PP such that $b_i\not\in I$ for all $i$. 
An interpretation $I$ is a \emph{stable model} of \PP if it is the $\subseteq$-minimal model of $\PP^I$ \cite{iclp/GelfondL88}.

In practice, often not just rules of the form \eqref{eq:rule}, but also aggregates are used. 
In non-ground programs (i.e., programs with first-order variables), they take various forms, but at the propositional level, it is well-known \cite{ai/SimonsNS02,tplp/MarekNT08} that in order to capture the standard aggregates~\mycite{AspCore2}, it suffices to consider only \emph{weight constraint rules}: rules of the form 
\begin{equation}\label{eq:rulePB} 
h \lrule W
\end{equation}
where $h\in\sigma$ and $W$ is a pseudo-Boolean
constraint $l \leq  \sum_i v_i a_i + \sum_i w_i\overline{b_i}$ with
$h$, the $a_i$'s, and the $b_i$'s in $\voc$ and with $l,v_i,w_i\in\mathbb{Z}$. 
Various semantics have been proposed for programs with weight constraint rules; for completeness we here include one. The \emph{FLP-reduct} of a program $\PP$ (with weight constraints) with respect to $\I$ is the set of rules of $\PP$ whose body is satisified in \I. A interpretation \I is an FLP-stable model of \PP if it is a minimal model of the FLP-reduct of \PP with respect to \I.  We do stress that the particular choice of semantics for these programs with weight constraints is not relevant in the current work, since we focus on the class of programs on which all proposals coincide. 
This is discussed in detail in the next section, and we come back to this issue in our discussion on future  work in Section \ref{sec:future}. 

\section{Translating Logic Programs into Pseudo-Boolean Theories}\label{sec:translation}

\paragraph{Scope and Limitations}
As can be seen in our definition of rules, we do not consider so-called  \emph{disjunctive} logic programs in this paper: the head of a rule is a single atom. 
For programs where disjunction ``behaves nicely'', for instance for the classes of head-cycle free \mycite{HCF-LP} and head-elementary-set-free \mycite{HEF-LP} programs, disjunction in the head can be eliminated by means of an operation called \emph{shifting} \cite{kr/GelfondPLT91}. Through the use of \lpshift \cite{ki/Janhunen18a}, which implements shifting, our tool also works for such programs. 

For weight constraint programs, or more generally, programs with different notions of aggregates, many different semantics have been proposed \cite{lpnmr/Ferraris05,corr/SonPE06,tplp/PelovDB07,ai/FaberPL11,tplp/GelfondZ14,ki/AlvianoF18}. 
Following the ASP-Core-2 standard~\mycite{AspCore2}, we restrict our attention to programs \emph{without recursion over aggregates} since for such programs all of the aforementioned semantics coincide.
Formally, we say that in the context of a logic program $\PP$, an atom $h$ \emph{depends directly on} atom $h'$ if there is a rule $r$ (of the form \eqref{eq:rule} or \eqref{eq:rulePB}) in \PP where $h'$ occurs in $\body(r)$. We say that $h$ \emph{depends} on atom $h'$ if it depends directly on $h'$ or if there exists some $h''$ such that $h$ depends on $h''$ and $h''$ on $h'$. 
In this paper, following the ASP-Core-2 standard, we only consider programs such that for each rule of the form \eqref{eq:rulePB}, no atom $h'$ that occurs in $W$ depends on $h$. 

\paragraph{Translation}
In order to translate logic programs into pseudo-Boolean theories, we make use of existing frameworks and tools as much as possible, to avoid reinventing the wheel. 
First of all, we can assume that for each rule of the form \eqref{eq:rulePB} in the program, $h$ is uniquely defined by that rule. 
We can always obtain this situation by introducing a new atom $h'$ and replacing the rule by two rules 
$ h'\lrule W$ and $h\lrule h'$. This operation is sound for most semantics of logic programs with  weight constraint rules, and it is always sound if there is no recursion over aggregates. 
It now becomes apparent that the aggregates can be ``isolated''. 
\begin{proposition}\label{prop:main}
Let \PP be a logic program without recursion over aggregates  such that for each weight constraint rule $r\in\PP$, $\head(r)$ has no other defining rules. Let $\PP'$ be the logic program obtained from $\PP$ by replacing all constraint rules $h\lrule W$ by two rules $h\lrule \overline{h'}$ and $h'\lrule \overline{h}$, where $h'$ is a new atom not occurring in \PP. Then there is a one-to-one correspondence between the answer sets of $\PP$ and the answer sets of $\PP'$ in which $h \Leftrightarrow W$ is satisfied for each rule of the form \eqref{eq:rulePB} in $\PP$. 
\end{proposition}
This (unsurprising) proposition follows directly from well-known splitting results; for instance the seminal work of Lifschitz and Turner~\cite{iclp/LifschitzT94} or the results of Vennekens et al.~\cite{tocl/VennekensGD06} for an algebraic variant that is applicable to the semantic characterization of Pelov et al.~\cite{tplp/PelovDB07} of logic programs with aggregates. 

Proposition \ref{prop:main} shows that we can split the task of translating $\PP$ into a pseudo-Boolean theory in two parts: first, we use any off-the-shelve method to translate $\PP'$ into a propositional theory, and next, we add an encoding of the constraints of the form $h \Leftrightarrow W$, for instance using constraints of the form
\[ h\Leftrightarrow b \leq  \sum_{i=1}^n w_i l_i\] 
is equivalent to the set of pseudo-Boolean constraints 
\begin{align*}
   b \leq  \sum_{i=1}^n w_i l_i  + M_1 \overline{h}, \qquad \qquad
   b >  \sum_{i=1}^n w_i l_i  - M_2 h
\end{align*}
when $M_1$ and $M_2$ are sufficiently large. 
The equivalence holds as soon as $M_1\geq b - \sum_{i=1}^n \min(0,w_i)$ and $M_2 > b + \sum_{i=1}^n \max(w_i,0)$. 
For instance, for such large $M_i$, in case $h$ is false, the first constraint is trivially satisfied; in case $h$ is true, it reduces to $
   b \leq  \sum_{i=1}^n w_i l_i.$

\section{Implementation}\label{sec:implementation}
Our tool accepts input in the \lparse--\smodels intermediate format \mycite{lparsemanual}. 
In case disjunction is present in the head of a rule, it is first eliminated using \lpshift~\cite{ki/Janhunen18a}. 
Of course, this is not correct in general, but only for the classes of programs considered here, where disjunction is head-cycle free (or, more general, head-elementary set free). 
The input is split into two parts: one part contains all rules containing aggregates (in the \lparse--\smodels format these are the constraint rules, weight rules, and minimize rules) while the other part contains all other rules, as well as the other information present in the \lparse--\smodels intermediate format (the symbol table, and compute statements). 
For constraint and weight rules (the former are a special case of the latter) in the first part, the transformation from Proposition~\ref{prop:main} is used to split them into a choice rule (the combination of the rules $ h\lrule\bar{h'}, h'\lrule\bar{h}$) which is added to the second part and two pseudo-Boolean constraints as described below Proposition~\ref{prop:main} to be included in the output. 
A minimize statement directly corresponds to a linear term, and is hence translated directly into a corresponding minimisation statement in the OPB format. 
As mentioned in Section~\ref{sec:translation}, we make use of existing tools as much as possible. Therefore, the second part (with the additional rules) is then given to the pipeline $ \lptonormal  \mid  \lptolp  \mid \lptosat$; the combination of these three tools translates a non-disjunctive logic program into an equivalent propositional theory in CNF \cite{lpnmr/Bomanson17,ki/Janhunen18a}. 
Our tool subsequently transforms each clause produced by \lptosat into a simple linear constraint and combines this with the linear constraints obtained from the first part to produce a complete pseudo-Boolean theory that characterizes exactly the stable models of the original logic program. 
We do not describe the complete implementation, but instead discuss a couple of peculiar points. 

\paragraph{Auxiliary Variables}
Since the translation introduces auxiliary variables, the translation happens only after parsing the entire input; at that point the highest used variable number is known; auxiliary variables will be numbered with subsequent numbers. 

\paragraph{Multilevel optimization}
While the ASP-Core-2 standard supports multilevel optimization (expressed in the \lparse--\smodels intermediate format by multiple minimise rules), the OPB format has no such construct. 
We use a well-known technique to reduce multilevel optimization to single-level optimization, namely summing up the different optimization terms but thereby multiplying the optimization terms at higher levels with a coefficient that is large enough to dominate over the terms at the lower levels. 
An effect of this is that --- without postprocessing of the results produced by the pseudo-Boolean solver --- the actual values of the optimization function cannot be read out directly from the output. 

\paragraph{The closed world assumption}
Answer set programming uses a form of the closed world assumption: all variables that are not mentioned in a program are false. 
In propositional logic on the other hand, unmentioned variables can take an arbitrary value. 
When naively applying the transformation from Proposition \ref{prop:main}, this can cause problems. 
For instance, if in the original program a certain variable only occurs in the body of a weight constraint rule (or in the optimization statement), then after applying the translation that variable no longer occurs in the program to be translated into SAT. 
Since in the original program, it is implicit that that variable must be false (due to the lack of any rules that derive it), it should still be false after translating. However, our pipeline used to translate to SAT does not enforce this constraint unless is it aware of the existence of that variable. There are two possible ways to fix this: either by including such variables in the symbol table or by manually adding a constraint that makes them false. We implemented the first option. 

\paragraph{Unused variables}
A last point of attention is that \lptosat, when translating a logic program into CNF makes some simplifications. In particular, in case a variable does not occur in the body of any rule, and that atom is not included in the symbol table (meaning that the user does not care about the value of that atom), \lptosat adds a constraint that falsifies this atom. 
However, since we only give \lptosat a part of the program, this optimization is no longer correct. 
This behaviour is again avoided by adding all atoms that occur in rules not given to the \lptosat pipeline in the symbol table.

\section{Experiments}\label{sec:experiments}
The experiments and set-ups were chosen to shine light on following research questions:


\begin{enumerate}
 \item How well do modern Pseudo-Boolean solvers perform on ASP models of problems where cutting planes is known to be stronger than resolution? 
 \item To which extent is the cutting plane proof system promising for traditional ASP?
\end{enumerate}

The benchmarks were ran on the VUB Hydra cluster. Each solver call was assigned a single core on a 10-core INTEL E5-2680v2 (IvyBridge) processor, a timelimit of 20 minutes and a memory-limit of 12GB, thereby matching the limits of the latest ASP competition \mycite{ASPComp7}.
The following benchmarks were used: 
\begin{enumerate}
 \item Four benchmark families inspired by the work of Elffers et al.~\cite{sat/ElffersGNV18}, using problems described there, as well as the same types of instances as in that paper (e.g., the shapes of the graphs considered).  These four families are known to be (with the right encoding) easy for the cutting-plane proof system in the sense that polynomial cutting plane proofs exist, but are hard for CDCL solvers. All our ASP encodings are straightforward and use aggregates. The four families are: 
 \begin{description}
  \item[Pigeon Hole] The problem here is to fit $n$ pigeons in $m$ holes with at most one pigeon residing in each hole. All our instances are unsatisfiable with $n=m+1$. 
  \item[Even Colouring] This problem takes as input a connected graph in which each vertex has an even degree. The problem is to determine if a black-white colouring of the edges exists such that each nodes has the same number of incident black and white edges. The problem is satisfiable if and only if the number of edges is even. Our instances are long toroidal grids in which one auxiliary vertex is inserted to break a single edge in two. All these instances are thus unsatisfiable. 
  \item[Vertex Cover] The input to this problem is a connected graph and a number $S$. The problem is to decide if a size $S$ vertex cover exists, i.e., a subset of the nodes of size $S$ such that each edge is incident to some vertex in the set. 
  We again use long toroidal grids, here with an even number of rows; in that case an instance is satisfiable if and only if $S\geq m\cdot \lceil n/2\rceil$ where $m$ is the number of rows and $n$ the number of columns. All our instances are unsatisfiable and have $S=m\cdot \lceil n/2\rceil -1$. 
  \item[Dominating Set] This problem again takes a graph and a number $S$ as input. The problem is to decide if the input graph has a size-$S$ dominating set, i.e., a set of vertices such that each vertex is either in the set or adjacent to a vertex in teh set. Our instances are long hexagonal grid. All our instances are unsatisfiable and have $S=\lfloor v/4\rfloor$ where $v$ is the number of vertices in the graph. 
 \end{description}
 The instances selected in these four benchmark families all scale linearly, that is, after starting from a small instance, we increase the size of the instance by a fixed step size. 
\item All decision and optimization problems from the 2017 ASP competition \mycite{ASPComp7}, which includes many benchmarks from earlier competitions, with the exception of: 
\begin{itemize}
 \item Problems including non head-cycle-free disjunction, since those were problems beyond the first level of the polynomial hierarchy that can hence not be translated compactly into pseudo-Boolean theories. 
 \item The video streaming benchmark family, since it contains very high coefficients (higher than what \roundingsat supports). 
\end{itemize}
For each benchmark family, the 20 instances selected for the competition were used. 
\end{enumerate}
All benchmarks and instances are available at \url{https://github.com/wulfdewolf/lp2pb_benchmarks}. 


\newcommand\configC{{\bf \m{\toolname{C}_\toolname{c}}}\xspace} 
\newcommand\configRS{{\bf \m{\toolname{C}_\toolname{pb}}}\xspace} 
\newcommand\configRSN{{\bf \m{\toolname{C}_\toolname{n-pb}}}\xspace} 
\newcommand\configCN{{\bf \m{\toolname{C}_\toolname{n-c}}}\xspace} 

We compared four solver configurations: 
\begin{itemize}
 \item \configC: $\gringo \mid \clasp$
 \item \configRS: $\gringo \mid \lptopb \mid \roundingsat$
 \item \configRSN: $\gringo \mid \lptonormal \mid \lptolp \mid \lptosat \mid \roundingsat$
 \item \configCN: $\gringo \mid \lptonormal \mid \lptolp \mid \clasp$
\end{itemize}
Of each of the used tools, the latest available version was used, i.e. \gringo 5.4.0, \clasp 3.3.5, \lptonormal 2.27, \lptolp 1.23, \lptosat 1.24, \lptopb 1.0\footnote{\url{https://github.com/wulfdewolf/lp2pb}}, and \roundingsat at commit fd464d43a\footnote{At the time of the writing, this commit has not been released yet. For reproducability, the binary can be found on our experiment github repository.}. 

A comparison between 
\configC and \configRS should give insights into how a state-of-the-art ASP solvers compares to a state-of-the-art pseudo-Boolean solver after our translation. 
Interpreting the results of \configRSN requires some care. 
For decision problems, in \configRSN, the input given to \roundingsat is a CNF. 
It is well-known that despite the fact that cutting planes can be exponentially more powerful than resolution, this power is not used by \emph{conflict-driven} pseudo-Boolean solvers on CNF input, where they essentially produce resolution proofs (see e.g., \cite{sat/VinyalsEGGN18}). 
For \emph{decision problems}, a comparison between \configRSN and \configCN should thus give an idea of the difference in engineering and optimizations between \roundingsat and \clasp. For \emph{optimization problems}, this comparison does not hold since bounds on the objective function that are added during branch-and-bound search are typically non-clausal. 
Finally, a comparison between 
%
\configRS and \configRSN should give an indication of how valuable the pseudo-Boolean constraints (coming from aggregates) are for \roundingsat, i.e., how much is gained by using our translation compared to a plain CNF translation. 

\begin{figure}[t]
\centering
\begin{subfigure}[b]{0.47\textwidth}
\centering
\begin{adjustbox}{width=0.9\linewidth}
\begin{tikzpicture}[y=.008cm, x=.108cm,font=\sffamily]

	\draw (0,0) -- coordinate (x axis mid) (100,0);
    	\draw (0,0) -- coordinate (y axis mid) (0,1200);
    	\foreach \x in {0,10,...,100}
     		\draw (\x,0) -- (\x,-3pt)
			node[anchor=north] {\x};
    	\foreach \y in {0,60,...,1200}
     		\draw (0,\y) -- (-3pt,\y) 
     			node[anchor=east] {\y}; 
     			
    \draw (0,1200) -- coordinate (x axis m) (100,1200);
    	\draw (100,0) -- coordinate (y axis m) (100,1200);
    	\foreach \x in {10,20,...,90}
     		\draw (\x,1200*0.225-1pt) -- (\x,1200*0.225+3pt)
			node[anchor=north] {};
    	\foreach \y in {60,120,...,1140}
     		\draw (100*0.108cm-3pt,\y) -- (100*0.108cm,\y) 
     			node[anchor=east] {}; 
    
	\node[below=0.8cm] at (x axis mid) {instance};
	\node[rotate=90, above=0.8cm] at (y axis mid) {time (s)};

	\draw plot[mark=*, mark options={fill=white, scale=2}] 
		file {plots/cactus/DominatingSet/data1.dat};
	\draw plot[mark=triangle*, mark options={fill=green, scale=2} ] 
		file {plots/cactus/DominatingSet/data2.dat};
	\draw plot[mark=diamond*, mark options={fill=red, scale=2}]
		file {plots/cactus/DominatingSet/data3.dat};
	\draw plot[mark=square*, mark options={fill=blue, scale=1.5}]
		file {plots/cactus/DominatingSet/data4.dat};  
    
	\begin{scope}[shift={(5,950)}] 
	\draw (0,0) -- 
		plot[mark=*, mark options={fill=white}] (0.25,0) -- (0.5,0) 
		node[right]{\configC};
	\draw[yshift=\baselineskip] (0,0) -- 
		plot[mark=triangle*, mark options={fill=green}] (0.25,0) -- (0.5,0)
		node[right]{\configRS};
	\draw[yshift=2\baselineskip] (0,0) -- 
		plot[mark=diamond*, mark options={fill=red}] (0.25,0) -- (0.5,0)
		node[right]{\configRSN};
	\draw[yshift=3\baselineskip] (0,0) -- 
		plot[mark=square*, mark options={fill=blue}] (0.25,0) -- (0.5,0)
		node[right]{\configCN};
	\end{scope}
\end{tikzpicture}
\end{adjustbox}

\caption{Dominating Set}\label{fig:dominatingset}
\end{subfigure}
%
\begin{subfigure}[b]{0.47\textwidth}
\centering
\begin{adjustbox}{width=0.9\linewidth}
\begin{tikzpicture}[y=.008cm, x=.0216cm,font=\sffamily]

	\draw (0,0) -- coordinate (x axis mid) (500,0);
    	\draw (0,0) -- coordinate (y axis mid) (0,1200);
    	\foreach \x in {0,50,...,500}
     		\draw (\x,0) -- (\x,-3pt) 
     			node[anchor=north] {\x}; 
    		
    	\foreach \y in {0,60,...,1200}
     		\draw (0,\y) -- (-3pt,\y) 
     			node[anchor=east] {\y};

    \draw (0,1200) -- coordinate (x axis m) (500,1200);
    	\draw (500,0) -- coordinate (y axis m) (500,1200);
    	
    	\foreach \x in {50,100,...,450}
     		\draw (\x,1200*0.225-1pt) -- (\x,1200*0.225+3pt)
     			node[anchor=east] {}; 
    		
    	\foreach \y in {60,120,...,1140}
     		\draw (500*0.0216cm-3pt,\y) -- (500*0.0216cm,\y) 
     			node[anchor=east] {}; 
    
	\node[below=0.8cm] at (x axis mid) {instance};
	\node[rotate=90, above=0.8cm] at (y axis mid) {time (s)};

	\draw plot[mark=*, mark options={fill=white, scale=2}] 
		file {plots/cactus/EvenColouring/data1.dat};
	\draw plot[mark=triangle*, mark options={fill=green, scale=2} ] 
		file {plots/cactus/EvenColouring/data2.dat};
	\draw plot[mark=diamond*, mark options={fill=red, scale=2}]
		file {plots/cactus/EvenColouring/data3.dat};
	\draw plot[mark=square*, mark options={fill=blue, scale=1.5}]
		file {plots/cactus/EvenColouring/data4.dat};  
    
	\begin{scope}[shift={(50,950)}] 
	\draw (0,0) -- 
		plot[mark=*, mark options={fill=white}] (0.25,0) -- (0.5,0) 
		node[right]{\configC};
	\draw[yshift=\baselineskip] (0,0) -- 
		plot[mark=triangle*, mark options={fill=green}] (0.25,0) -- (0.5,0)
		node[right]{\configRS};
	\draw[yshift=2\baselineskip] (0,0) -- 
		plot[mark=diamond*, mark options={fill=red}] (0.25,0) -- (0.5,0)
		node[right]{\configRSN};
	\draw[yshift=3\baselineskip] (0,0) -- 
		plot[mark=square*, mark options={fill=blue}] (0.25,0) -- (0.5,0)
		node[right]{\configCN};
	\end{scope}
\end{tikzpicture}
\end{adjustbox}

\caption{Even Colouring}\label{fig:evencolouring}
\end{subfigure}
%
\begin{subfigure}[b]{0.47\textwidth}
\centering
\begin{adjustbox}{width=0.9\linewidth}
\begin{tikzpicture}[y=.008cm, x=.108cm,font=\sffamily]

	\draw (0,0) -- coordinate (x axis mid) (100,0);
    	\draw (0,0) -- coordinate (y axis mid) (0,1200);
    	\foreach \x in {0,10,...,100}
     		\draw (\x,0) -- (\x,-3pt)
			node[anchor=north] {\x};
    	\foreach \y in {0,60,...,1200}
     		\draw (0,\y) -- (-3pt,\y) 
     			node[anchor=east] {\y}; 
     			
    \draw (0,1200) -- coordinate (x axis m) (100,1200);
    	\draw (100,0) -- coordinate (y axis m) (100,1200);
    	\foreach \x in {10,20,...,90}
     		\draw (\x,1200*0.225-1pt) -- (\x,1200*0.225+3pt)
			node[anchor=north] {};
    	\foreach \y in {60,120,...,1140}
     		\draw (100*0.108cm-3pt,\y) -- (100*0.108cm,\y) 
     			node[anchor=east] {}; 
    
	\node[below=0.8cm] at (x axis mid) {instance};
	\node[rotate=90, above=0.8cm] at (y axis mid) {time (s)};

	\draw plot[mark=*, mark options={fill=white, scale=2}] 
		file {plots/cactus/PigeonHole/data1.dat};
	\draw plot[mark=triangle*, mark options={fill=green, scale=2} ] 
		file {plots/cactus/PigeonHole/data2.dat};
	\draw plot[mark=diamond*, mark options={fill=red, scale=2}]
		file {plots/cactus/PigeonHole/data3.dat};
	\draw plot[mark=square*, mark options={fill=blue, scale=1.5}]
		file {plots/cactus/PigeonHole/data4.dat};  
    
	\begin{scope}[shift={(5,950)}] 
	\draw (0,0) -- 
		plot[mark=*, mark options={fill=white}] (0.25,0) -- (0.5,0) 
		node[right]{\configC};
	\draw[yshift=\baselineskip] (0,0) -- 
		plot[mark=triangle*, mark options={fill=green}] (0.25,0) -- (0.5,0)
		node[right]{\configRS};
	\draw[yshift=2\baselineskip] (0,0) -- 
		plot[mark=diamond*, mark options={fill=red}] (0.25,0) -- (0.5,0)
		node[right]{\configRSN};
	\draw[yshift=3\baselineskip] (0,0) -- 
		plot[mark=square*, mark options={fill=blue}] (0.25,0) -- (0.5,0)
		node[right]{\configCN};
	\end{scope}
\end{tikzpicture}
\end{adjustbox}

\caption{Pigeon Hole}\label{fig:pigeonhole}
\end{subfigure}
%
\begin{subfigure}[b]{0.47\textwidth}
\centering
\begin{adjustbox}{width=0.9\linewidth}
\begin{tikzpicture}[y=.008cm, x=0.09cm,font=\sffamily]

	\draw (0,0) -- coordinate (x axis mid) (120,0);
    	\draw (0,0) -- coordinate (y axis mid) (0,1200);
    	\foreach \x in {0,10,...,120}
     		\draw (\x,0) -- (\x,-3pt)
			node[anchor=north] {\x};
    	\foreach \y in {0,60,...,1200}
     		\draw (0,\y) -- (-3pt,\y) 
     			node[anchor=east] {\y}; 
     			
    \draw (0,1200) -- coordinate (x axis m) (120,1200);
    	\draw (120,0) -- coordinate (y axis m) (120,1200);
    	\foreach \x in {10,20,...,110}
     		\draw (\x,1200*0.225-1pt) -- (\x,1200*0.225+3pt)
			node[anchor=north] {};
    	\foreach \y in {60,120,...,1140}
     		\draw (90*3.38-1pt,\y) -- (90*3.38+3pt,\y) 
     			node[anchor=east] {}; 
    
	\node[below=0.8cm] at (x axis mid) {instance};
	\node[rotate=90, above=0.8cm] at (y axis mid) {time (s)};

	\draw plot[mark=*, mark options={fill=white, scale=2}] 
		file {plots/cactus/VertexCover/data1.dat};
	\draw plot[mark=triangle*, mark options={fill=green, scale=2} ] 
		file {plots/cactus/VertexCover/data2.dat};
	\draw plot[mark=diamond*, mark options={fill=red, scale=2}]
		file {plots/cactus/VertexCover/data3.dat};
	\draw plot[mark=square*, mark options={fill=blue, scale=1.5}]
		file {plots/cactus/VertexCover/data4.dat};  
    
	\begin{scope}[shift={(5,950)}] 
	\draw (0,0) -- 
		plot[mark=*, mark options={fill=white}] (0.25,0) -- (0.5,0) 
		node[right]{\configC};
	\draw[yshift=\baselineskip] (0,0) -- 
		plot[mark=triangle*, mark options={fill=green}] (0.25,0) -- (0.5,0)
		node[right]{\configRS};
	\draw[yshift=2\baselineskip] (0,0) -- 
		plot[mark=diamond*, mark options={fill=red}] (0.25,0) -- (0.5,0)
		node[right]{\configRSN};
	\draw[yshift=3\baselineskip] (0,0) -- 
		plot[mark=square*, mark options={fill=blue}] (0.25,0) -- (0.5,0)
		node[right]{\configCN};
	\end{scope}
\end{tikzpicture}
\end{adjustbox}

\caption{Vertex Cover}\label{fig:vertexcover}
\end{subfigure}

\caption{Cactus plots for the first set of benchmark families.}
\label{fig:cactus}
\end{figure}
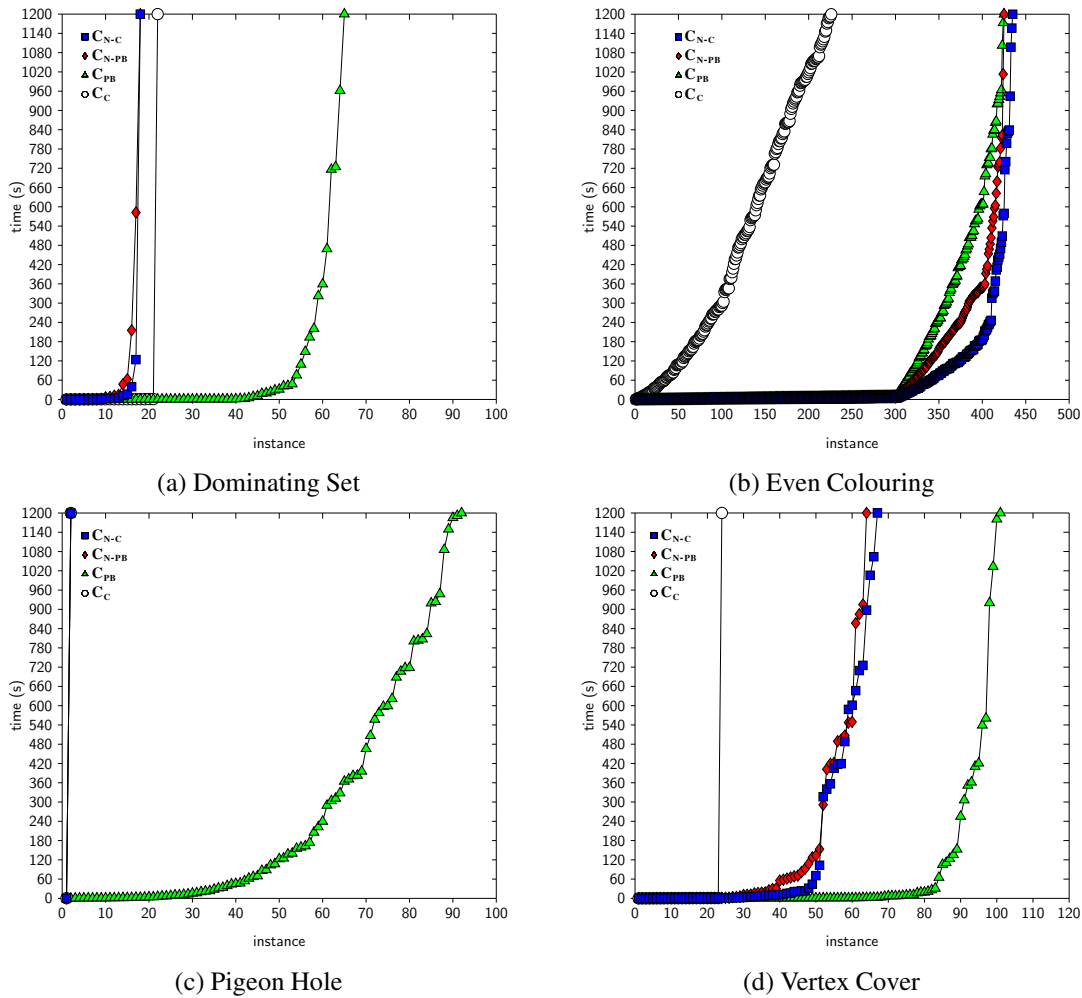

\paragraph{Analysis}
Cactus plots of the runtimes of the first benchmark set are presented in Figure~\ref{fig:cactus}. Overall, these results are consistent with our expectations. The combination of \lptopb and \roundingsat outperforms resolution-based solvers by far. 
This is most prominently visible in the Pigeon Hole problem, where no resolution-based configuration solves the problem with 16 pigeons, while \roundingsat solves all problems up 916 pigeons.
The odd one out of the four families is the Even Colouring family, where the normalization-based configurations slightly outperform \configRS. Our assumption is that the auxiliary variables introduced by $\lptonormal$ change the language of learning and in this way enable short resolution proofs. 
A similar effect, but less prominent, is seen in the Vertex Cover family, where normalization-based approaches also outperform \configC, but do not reach the performance of \configRS. 

%
%

\begin{table}[t]
\caption{\emph{Decision problems of the ASP competition. For each family and set-up pair, 20 instances were ran; this table contains the number of instances for which (un)satisfiability was proven.}}
\label{tab:decision}
\centering
\begin{tabular}{l@{\quad}ccccccc}
&&& \multicolumn{4}{c}{\textbf{(UN)SAT Proven}}\\
\textbf{Family}  & \textbf{\#sum?} & \textbf{\#count?} & \bf \configC & \bf \configRS &\bf  \configRSN & \bf \configCN \\ \hline\hline
Crew Allocation & No & Yes &\bf 18& 17& 15& 16 \\ 
Graph Colouring & No & No &\bf 16& 8& 8&\bf 16 \\ 
Knight Tour With Holes  & No & No &\bf 13& 2& 2&	3 \\ 
Labyrinth & No & No &\bf 13& 1& 3&	12 \\ 
Stable Marriage & No & No &\bf 8& 0& 0& 3 \\ 
Visit-all & No & Yes &\bf 18&	13&	13&	17 \\ 
Combined Configuration & Yes  & Yes &\bf 14& 2& 2& 1 \\ 
Graceful Graphs & No & Yes &\bf 13& 8&	10&\bf 13 \\ 
Incremental Scheduling  & Yes & Yes &\bf 14& 13& 1& 1 \\ 
Nomistery & No & No & 7&\bf 9& 7& 8 \\ 
Partner Units & No & Yes &\bf 11& 10& 10& 10 \\ 
Permutation Patternmatching & No & No &\bf 13& 9&	8& 8 \\ 
Qualitative Spatial Reasoning & No & No & 11& 11&\bf 13& 12 \\ 
Ricochet Robots & No & Yes & 10& 7&\bf 11&	10 \\ 
Sokoban & No & Yes &\bf 11& 9& 8& 10 \\ 
\hline
\bf Total &&&\bf 190& 119& 111& 140
\end{tabular}
\end{table}

When examining the results on decision problems, summarized in Table \ref{tab:decision}, we notice that \configC, i.e. $\gringo \mid \clasp$, outperforms all other configurations on most benchmark families. For problems without aggregates, this is in line of the expectations. But for problems with aggregates our expectation was to see a positive effect of the cutting planes proof system. 
Also the difference between \configRS and \configRSN is very small, suggesting that little to no benefit of the cutting plane proof system is obtained on those benchmarks. 

\tabcolsep=2pt
\begin{table}[t]
\caption{\emph{Optimization problems. For each family and configuration, the number of instances (out of 20) for which optimality was proven, as well as the number of instances for which a configuration found the best optimization value, among the four configurations. For the System Synthesis family this column cannot be calculated as the problems in this family are multilevel optimisation problems.}}
\label{tab:optimization}
\resizebox{\textwidth}{!}{
\begin{tabular}{l@{\quad}ccccccccccc}
&&& \multicolumn{4}{c}{\textbf{Optimality Proven}} && \multicolumn{4}{c}{\textbf{Best Value Found}}\\
\textbf{Family} & \textbf{\#sum?} & \textbf{\#count?} &  \bf \configC & \bf \configRS &\bf  \configRSN & \bf \configCN &\qquad & \bf \configC & \bf \configRS &\bf  \configRSN & \bf \configCN \\ \hline\hline
Bayesian NL & No & Yes & \bf 15& 5& 4 & 14 &&\bf 18 & 10 & 9 &\bf 18 \\ 
Markov NL & No & Yes & \bf 11& 0& 0& 9 && \bf 18 & 5 & 4 & 16 \\ 
Supertree & No & Yes & 7& 5& 5&\bf 8 && 13 & 5 & 5 & \bf 17 \\ 
Connected Maximum-density Still Life & No & Yes & 7&\bf 8&\bf 8& 2 && \bf 19 & 8 & 9 & 7 \\ 
Crossing Minimization & No & Yes &13&\bf 19&\bf 19&13&  & 15 & \bf 20 & 19 & 17 \\ 
Maximal Clique Problem & No & No & 0& 0& 0& 0 && 0 & \bf 16 & 13 & 0 \\ 
Max SAT & No & Yes & 10&\bf 18&\bf 18& 10 && 11 & \bf 19 & \bf 19 & 11 \\ 
Steiner Tree & No & Yes &\bf 3& 1& 1& 2 && \bf 20 & 2 & 2 & 3 \\ 
System Synthesis & Yes & Yes &0&0&0&0& & - & - & - & - \\ 
Valves Location problem & Yes & Yes &\bf 15& 13& 3& 6 && \bf 20 & 13 & 3 & 6 \\ 
\hline
\bf Total &&&\bf 81& 69& 58& 64 &&\bf 170& 120& 108& 110
\end{tabular}}
\end{table}

The optimization problems paint a different picture. Next to the number of instances completely solved for each family, Table \ref{tab:optimization} also shows the number of instances for which a given configuration found the best solution among the four configurations. 
For System Synthesis, those last values are not included as this is a multilevel optimization problem and the values given by the different solvers are incomparable. 
In three out of ten families a clear improvement of cutting plane over state-of-the-art ASP solving is visible. 
Overall, \configC is still the best solving configuration, but \configRS comes second, performing slightly better than the approaches in which aggregates are normalized. 
These results suggest that \lptopb, constitutes a valuable  extra tool in the ASP toolkit.


\section{Related Work}\label{sec:related}
There is a rich history of research on \textbf{using SAT solvers to search for computing stable models} of logic programs. One approach works by introducing loop formulas on-the-fly \cite{ai/LinZ04,lpnmr/Lierler05,jar/GiunchigliaLM06} and in fact lies at the basis of most modern native ASP solvers \cite{\refto{clasp},\refto{wasp}}. 
Another approach is studying \textbf{translations of ASP into SAT} that are more compact than the (worst-case) exponential blow-up loop formulas induce  \cite{amai/Ben-EliyahuD94,ijcai/LinZ03,ecai/Janhunen04,birthday/JanhunenN11}. These translations introduce auxiliary variables; some of them induce a one-to-one correpondence between the stable models of the original program and the models of the obtained propositional theory, while others duplicate some models. 
What all these methods have in common is that aggregates are encoded into clause, either lazily (as done in native solvers, often following the lazy clause generation paradigm \mycite{LCG}), or eagerly, for instance by applying normalization tools \cite{lpnmr/Bomanson17} that eliminate the aggregates before the actual SAT-translation is called.
Our work closely relates to the translation based approach. In fact, internally our tool makes use of the tools of Janhunen and Niemel{\"a}~\cite{birthday/JanhunenN11} to translate the part of the program without aggregates into SAT; the actual translation used can easily be changed in \lptopb. The main difference with the standard translational approaches is that aggregates are not normalized but preserved. 

Also \dlvtwo \cite{DBLP:conf/lpnmr/AlvianoADLMR19} has an option (\verb+--pre=wbo+) to translate ASP programs into PB theories; however, this translation is limited to tight programs and it cannot handle multilevel optimisation problems.

Another related tool is \mingo \mycite{mingo}, which integrates answer set programming and \emph{mixed integer programming} \mycite{MIP}, thus allowing more types of constraints (using non-Boolean variables) than PB theories. 
These non-Boolean variables are used, among others, to encode the level mapping characterization \cite{fuin/LiuY10} that ensures stability of the obtained models. 
Unlike our translation, \mingo does not guarantee a one-to-one correspondence between the models of the obtained theory and the stable models of the original program, making it unsuitable for model counting. Our approach does guarantee this, mainly by building on the guarantees from the used translation of aggregate-free logic programs to SAT~\cite{birthday/JanhunenN11}.
Another difference is that part of the focus of \mingo is on developing an extension of the ASP language in which mixed-integer constraints can be written directly in the program. Nowadays, this approach is common in various Constraint-ASP formalisms~\cite{aaai/DrescherW11,aaai/Lierler12,tplp/BanbaraKOS17,tplp/JanhunenKOSWS17,kr/ShenL18}.

A final related tool is \pbmodels \mycite{pbmodels}, which also uses \textbf{pseudo-Boolean solvers} to find stable models. The main difference is that \pbmodels is designed as a wrapper around a PB solver that iteratively calls the solver for supported models, next checks for stability and if the result is not stable, adds loop formulas, while \lptopb outputs a translation that can be fed to a PB solver to be solved in a single solver call, which benefits the solver's internal constraint learning mechanism.


\section{Conclusion and Future Work}\label{sec:conclusion}\label{sec:future}
One direction for future work is investigating an extension of our translation to support recursive aggregates. 
The semantics of recursive aggregates constitute an intense topic of debate, as can be witnessed by the number of papers written about them \cite{lpnmr/Ferraris05,corr/SonPE06,tplp/PelovDB07,ai/FaberPL11,tplp/GelfondZ14,ki/AlvianoF18}. 
However, for monotonic (and in fact, \emph{convex} aggregates \cite{jair/LiuT06}), most of them agree  --- the notable exception being \cite{tplp/GelfondZ14}. 
Hence, an extension of our tool that works for recursive aggregates under the condition that they be convex, would be valuable. 
The most lightweight way to achieve this would be to start from the translation of \mycite{mingo}, which builds on the level mapping of Liu and You~\cite{fuin/LiuY10}, and modify it to use Boolean variables. 
An unresolved challenge in that case is how a one-to-one correspondence between the stable models of the program and the models of the resulting theory can be achieved. 
%
%

Another interesting, but perhaps more ambitious direction for future work is to develop a new native ASP solver that uses the cutting plane proof system under the hood, for instance by developing an extension of \roundingsat with support for recursive rules. 


To conclude, we presented a novel tool, called \lptopb, to translate logic programs into pseudo-Boolean formulas and experimentally validated its performance on a large set of benchmarks. The results are mixed. On the one hand, overall traditional ASP solvers seem to outperform pseudo-Boolean solvers on the benchmark traditionally tackled with ASP. But on the other hand, a couple of benchmark families was identified on which pseudo-Boolean reasoning can provide a real advantage, thus warranting further research into using the cutting plane proof system in ASP solving. 

\section*{Acknowledgments}
The resources and services used in this work were provided by the VSC (Flemish Supercomputer Center), funded by the Research Foundation - Flanders (FWO) and the Flemish Government.
We are very grateful to Jakob Nordstr\"om and Jo Devriendt for interesting discussions on this topic and for providing us with the latest version of \roundingsat.

\bibliographystyle{eptcs}
\bibliography{idp-latex/krrlib,more}


\label{lastpage}
\end{document}
